\documentclass[%
reprint,
showpacs,
bibnotes,
 amsmath,amssymb,
 aps,
prb,
]{revtex4-1}

\usepackage{graphicx}
\usepackage{dcolumn}
\usepackage{bm}

\usepackage{epsfig}
\usepackage{mathrsfs}
\usepackage{amssymb}
\usepackage{amsbsy}
\usepackage{color}
\usepackage{cancel}
\newcommand{\bcen}{\begin{center}}
\newcommand{\ecen}{\end{center}}
\newcommand{\btab}{\begin{tabular}}
\newcommand{\etab}{\end{tabular}}
\newcommand{\bdes}{\begin{description}}
\newcommand{\edes}{\end{description}}

\newcommand{\beq}{\begin{equation}}
\newcommand{\eeq}{\end{equation}}
\newcommand{\bea}{\begin{eqnarray}}
\newcommand{\eea}{\end{eqnarray}}
\newcommand{\non}{\nonumber}

\newcommand{\half}{\frac{1}{2}}
\newcommand{\bary}{\begin{array}}
\newcommand{\eary}{\end{array}}
\newcommand{\benum}{\begin{enumerate}}
\newcommand{\eenum}{\end{enumerate}}
\newcommand{\bitem}{\begin{itemize}}
\newcommand{\eitem}{\end{itemize}}

\newcommand{\bsig}{\mbox{\boldmath $ \sigma $}}

\newcommand{\be} { \mbox{\boldmath $e$}}

\newcommand{\bk} { \bm{k} }
\newcommand{\br} { \boldsymbol{r}}

\newcommand{\bE} { \mbox{\boldmath $E$}}

\newcommand{\bK} { \boldsymbol{K}}

\newcommand{\bP} { \mbox{\boldmath $P$}}
\newcommand{\bQ} { \boldsymbol{Q} }

\newcommand{\bzero} { \mbox{\boldmath $0$}}

\newcommand{\D}[1]{\mbox{d}{#1}} 
\newcommand{\grad}{\mbox{\boldmath $\nabla$}}

\newcommand{\mean}[1]{\langle #1 \rangle}
\newcommand{\bra}[1]{{\langle #1 |}}
\newcommand{\ket}[1]{| #1 \rangle}

\newcommand{\prn}[1] {(\ref{#1})}

\newcommand{\fig}[1]{fig.~\ref{#1}}
\newcommand{\Fig}[1]{Fig.~\ref{#1}}

\newcommand{\myfigwidth}{0.99\columnwidth}
\newcommand{\signum}[0]{\mathop{\mathrm{sgn}}}

\begin{document}


\title{Sensory Organ like Response of Zigzag Edge Graphene Nanoribbons}

\author{Somnath Bhowmick}
 \email{bsomnath@mrc.iisc.ernet.in}
 \affiliation{Materials Research Centre, Indian Institute of Science, Bangalore 560 012}
\author{Vijay B.~Shenoy}%
\email{shenoy@physics.iisc.ernet.in}
\affiliation{Centre for Condensed Matter Theory, Indian Institute of Science, Bangalore 560 012}



\date{\today}

\begin{abstract}

Using a continuum Dirac theory, we study the density and spin response
of zigzag edge terminated graphene ribbons subjected to edge
potentials and Zeeman fields. Our analytical calculations of the
density and spin responses of the closed system (fixed particle
number) to the static edge fields, show a highly nonlinear
Weber-Fechner type behavior where the response depends
logarithmically on the edge potential.  The dependence of the response
on the size of the system (e.g.~width of a nanoribbon) is also
uncovered. Zigzag edge graphene nanoribbons, therefore, provide a
realization of response of organs such as the eye and ear that obey
Weber-Fechner law. We validate our analytical results with tight
binding calculations. These results are crucial in understanding
important effects of electron-electron interactions in graphene
nanoribbons such as edge magnetism etc., and also suggest
possibilities for device applications of graphene nanoribbons.
\end{abstract}


\maketitle

\section{Introduction}

Since it's first isolation\cite{novoselov2004}, graphene, a single
layer of carbon atoms arranged on a honeycomb lattice, has attracted
much attention.  Several unconventional and novel phenomena have been
observed in graphene; these include quantum Hall effect at room
temperature, Klein paradox etc., to name a
few.\cite{geim2007,katsnelson2007,neto2007} Many of its interesting
properties originate from the fact that energy spectrum of graphene, a
zero band-gap semiconductor with linearly vanishing density of
states(DOS) near the chemical potential, resembles the Dirac spectrum
of massless Fermions.\cite{semenoff1984} Due to it's unusual
electronic properties, graphene holds promise of replacing
semiconductor based transistors in future nanoelectronic and
spintronic devices.\cite{westervelt2008} Advent of several electronic
devices, such as single electron transistor based on graphene
flakes\cite{bunch2005} and nanodots of size as small as $\sim$ 30
nm,\cite{ponomarenko2008} quantum interference devices\cite{miao2007,Wurm2009}
etc.~are just the beginning of a fascinating future.

Several interesting nanostructures can be fabricated from
graphene. The simplest of them is the quasi one-dimensional nanoribbon,
made by terminating the graphene sheet by two edges of either
``armchair'' or ``zigzag'' type.\cite{nakada1996} Such terminations can
significantly influence the electronic structure of the graphene nanoribbon. In
particular, tight binding analysis of graphene bound by two zigzag
edges shows that there are localized electronic states at the edge
with nearly flat energy dispersion.\cite{nakada1996} In terms of
magnetic structure, half filled zigzag edge nanoribbons are
reported\cite{okada2001} to have antiferromagnetic ground state, with
very large moments localized at the edge sites compared to bulk and
the magnetizations of the opposite edges being antiferromagnetically aligned. The
investigation has further been extended to zero dimensional graphene
nanodots (half filled or undoped) of different shapes such as
rectangular, hexagonal etc.~\cite{jiang2007,rossier2007}, which are
also reported to be antiferromagnetic (zero net magnetization) above a
critical size.  There has also been interest in valley physics\cite{Rycerz2007,Abergel2009}, giant magnetoresistance\cite{Zhang2010, Bai2010}  phenomena in graphene nanoribbons. These developments strongly motivate studies aimed at
understanding the electronic properties of graphene nanosystems and
study of their response to stimulus.

Responses of quantum systems, including nanosystems, are usually linear
in the stimulus, with response functions described by Kubo
formulae.\cite{Altland2006} Nature also provides realization of
systems that inherently have a strongly nonlinear response. Key examples
of these are the eyes and ears of living organisms whose response is
described by the Weber-Fechner law.\footnote{See, for example, http://en.wikipedia.org/wiki/Weber-Fechner\_law} If $S$
is the stimulus and $R$ is the response, a system obeying the
Weber-Fechner law satisfies 
\bea 
\Delta R \sim \frac{\Delta S}{S} 
\eea
where $\Delta R$ is the change in the response for a change $\Delta S$
of the stimulus. This implies that the response depends
logarithmically on the stimulus. Indeed, this is why eyes and ears of
living organisms can sense light and sound of intensities that range
over 10--15 decades! Indeed, it is interesting to explore quantum/nano-systems that have such a nonlinear response.

In this paper we demonstrate theoretically that zigzag edge terminated
graphene nanoribbons subjected to edge potentials show responses that
are highly nonlinear and, most interestingly, obey the Weber-Fechner law. Our
motivation for this work arose from the earlier works on edge state
magnetism cited above that we found to be very
robust.\cite{bhowmick2008} Edge potentials can therefore be applied
externally or can be ``self generated'' by the system due to
interactions (as in the case of edge state magnetism), and the
considerations of this paper apply to both cases. Specifically, we
analytically solve, using a continuum Dirac theory for graphene, for
responses such as particle density and spin density (of the closed system with fixed number of particles) to applied static 
edge potentials and Zeeman fields in zigzag edge terminated graphene
nanoribbons. We show that they have a Weber-Fechner type nonlinear
response that can be tuned using the size (width) of the
ribbon  (see eqns. \prn{eqn:sym_M} and \prn{eqn:asym_M}). Indeed this work provides an uncommon example of a quantum
system whose nonlinear response can be calculated analytically, in addition to suggesting many novel uses of zigzag terminated graphene nanoribbons. We also compare our analytical calculations with full tight binding simulations and demonstrate excellent agreement between the two.

The paper is organized as follows. Section \ref{GNRReview} reviews the continuum field theory of graphene and the earlier work on the electronic structure of zigzag terminated nanoribbons, and in addition introduces the notation used in the paper. In Section \ref{sec:GNREdgePotential} we state and solve the problem of the zigzag terminated graphene nanoribbons subjected to edge potentials. This is followed by sections \ref{sec:dens_resp} and \ref{sec:spin_resp} where the density and spin responses to the edge potential are discussed and shown to have Weber-Fechner type behavior. These analytical results are shown to be in excellent agreement with full tight binding calculations in Section \ref{sec:compare_tb}. The paper is concluded in Section \ref{sec:summary} which includes a discussion of possible applications and future directions of research.

\section{Graphene Nanoribbons}
\label{GNRReview}

\begin{figure}
\centerline{\epsfxsize=\myfigwidth \epsfclipon \epsfbox[100 360 496 716]{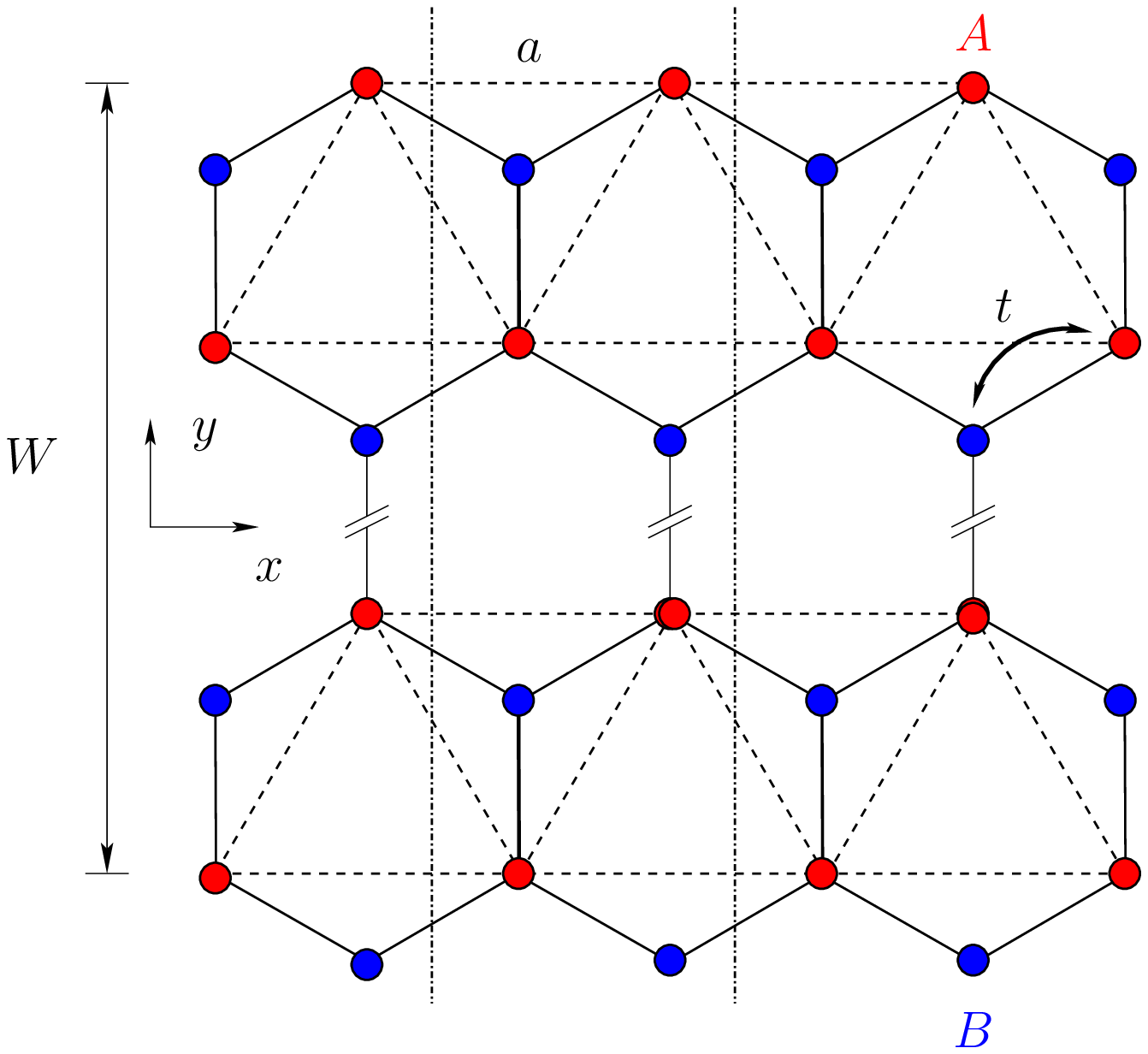}}
\caption{(color online) Graphene structure/zigzag edge terminated nanoribbon-geometry explaining notation used in the text. The triangular lattice parameter and the ``unit cell width'' of the nanoribbon is denoted by $a$; the atoms enclosed by the two dashed lines form the repeat unit basis along the $x$-direction. $W$ is the width of the nanoribbon.  The two sublattices $A$ and $B$ are color coded in red and blue respectively. The hopping amplitude for electrons is denoted by $t$.  }
\label{grscheme}
\end{figure}

The purpose of this section is two fold: to present a brief review of the electronic structure of graphene and graphene nanoribbons (for a detailed review cf.~\cite{neto2007}), and in the process introduce the notation used in the text. 

The low energy electronic degrees of freedom in graphene are the $p_z$ orbitals on each carbon atom. Electron hopping, with amplitude $t$, from $p_z$ orbital of one carbon atom to one of its three neighbors in a honeycomb lattice (see \Fig{grscheme}) is described by a tight binding model
\bea
{\cal H}_{TB} = -t \sum_{\mean{ij}} c^{\dagger}_{i \sigma} c_{j \sigma} \label{HTB}
\eea
where $c^\dagger_{i \sigma} (c_{i \sigma})$ creates (destroys) an electron of spin $\sigma$ at lattice site $i$. As is well known\cite{neto2007}, diagonalization of this hamiltonian leads to two bands which ``touch'' at two non-equivalent points $\bK_+$ and $\bK_{-}$ in the first Brillouin zone of the honeycomb lattice -- called the Dirac points -- that are related by time-reversal symmetry. By particle-hole symmetry of the hamiltonian, the energy where the bands touch is zero.

\begin{figure}
\centerline{\epsfxsize=\myfigwidth \epsfclipon \epsfbox[15 312 593 721]{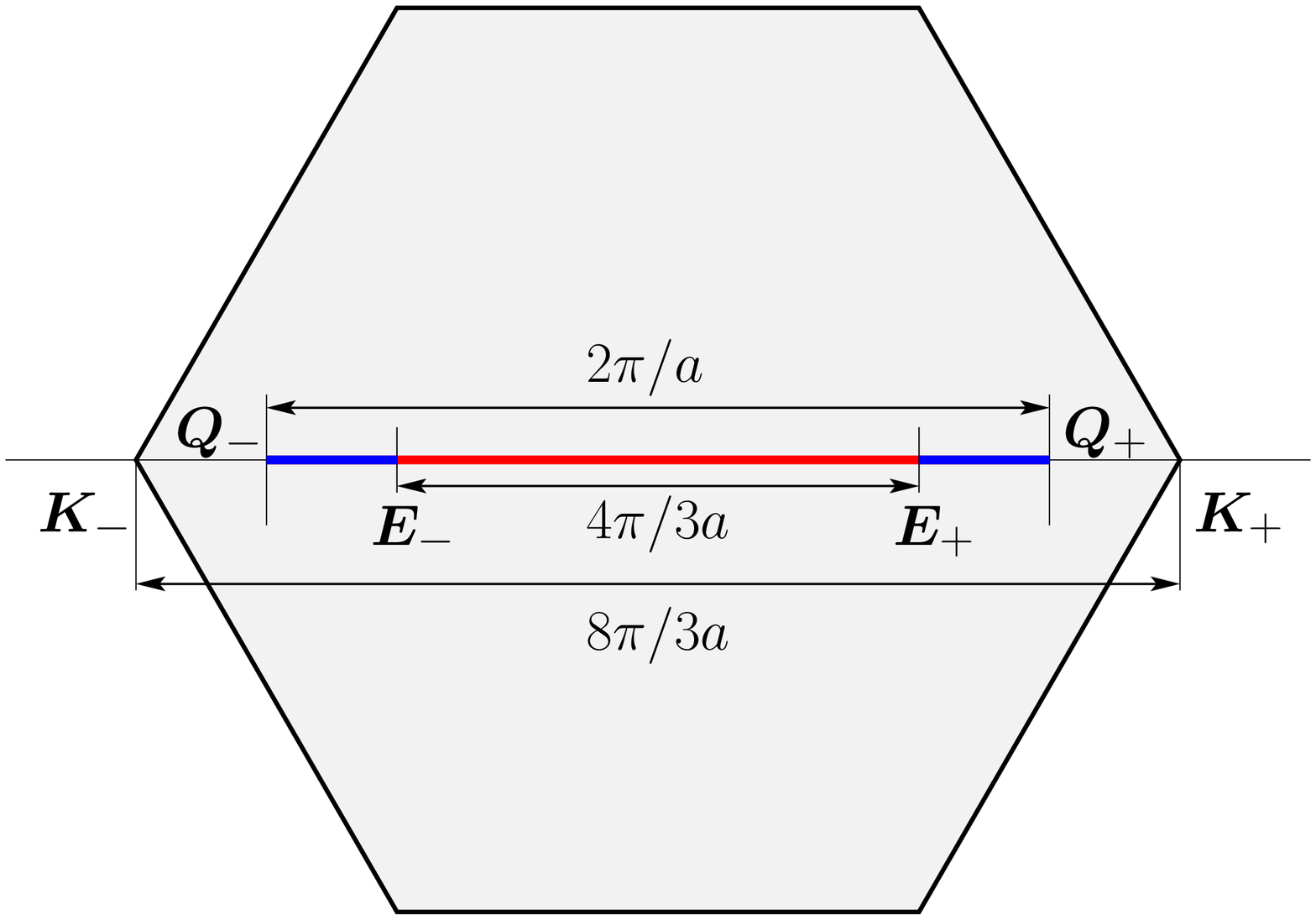}}
\caption{(color online) Brillouin zone of graphene. The points $\bK_{\pm} = (\pm \frac{4 \pi}{3 a}, 0)$ are the Dirac points. The line segment $\bQ_{-}\bQ_{+}$ represents the Brillouin zone of a zigzag nanoribbon (NRBZ) (segments in blue and red) with $\bQ_{\pm} = (\pm \frac{\pi}{a},0)$. Segments $\bE_{+}\bQ_+$ and $\bQ_{-}\bE_{-}$ corresponds to ``edge Brillouin zone'' (EBZ) (segments in blue) and is one third of the NRBZ ($\bE_{\pm} = (\pm \frac{2 \pi}{3 a}, 0)$). For the nanoribbon, the Dirac point ${\bK}_-$ is equivalent to $\bE_+$, and $\bK_+$ to  $\bE_-$.  }
\label{ebz}
\end{figure}

Each carbon atom contributes one $p_z$ electron, so the chemical potential in undoped graphene is zero and the highest occupied states are at the Dirac points. For this case, and for lightly doped graphene, the electronic excitations can be conveniently described by a continuum theory. The electronic low energy excitations (for each spin $\sigma$) are described  by two component wavefunction $\psi_{\sigma \alpha}(\br)$, where $\alpha$ denotes the ``sublattice flavor'' and can be $A$ or $B$ (see \Fig{grscheme}). Moreover, since the Bloch wavefunctions that contribute to these wavefunctions mainly arise from the neighborhood of ${\bK}_+$ and ${\bK}_-$ points (called as the $+$ and $-$ valleys), the wavefunction can be expanded as
\bea
\psi_{\sigma \alpha}(\br) = \psi_{\sigma \alpha +}(\br) e^{i {\bK}_+ \cdot \br} + \psi_{\sigma \alpha -}(\br) e^{i {\bK}_- \cdot \br} 
\eea
where $\psi_{\sigma \alpha v}(\br)$ ($v = \pm$), stands for wavefunctions with ``valley flavor''. For each spin, therefore, the wavefunction is described by a four component spinor, $\Psi_\sigma(\br) = \left(\psi_{\sigma A +}(\br)\;\;\psi_{\sigma B +}(\br)\;\;\psi_{\sigma A -}(\br)\;\;\psi_{\sigma B -}(\br) \right)^T$, which satisfies the Schr\"odinger equation
\bea
{\cal H_K} \Psi_\sigma = E \Psi_\sigma \label{Schrod}
\eea
with $E$ the energy eigenvalue, where 
\bea
{\cal H}_{K} = \hbar v_F \left( 
\begin{array}{cc}
\bsig \cdot \bP  & \bzero \\
\bzero & -\bsig^* \cdot \bP \\
\end{array}
\right) \label{HK}
\eea
where $v_F = \frac{e_0 a}{\hbar}$ is the Fermi velocity at the Dirac points, $\bP = - i \hbar \grad$ and $\bsig = \be_x \sigma_x + \be_y \sigma_y$; $\sigma_{x,y}$ are the Pauli matrices. The energy $e_0$ is related to the hopping parameter via $e_0 = \frac{\sqrt{3} t}{2}$. The hamiltonian has an emergent SU(4) symmetry.\cite{Herbut2006}  The eigenstates of the hamiltonian \prn{HK} are linearly dispersing modes with energy $\hbar v_F |\bk|$ about each of the Dirac valleys.

Consider, now, an external potential $V_{\alpha}(\br)$ that acts on the lattice. Assuming that the potential is ``slowly varying'', the Hamiltonian becomes
\bea
{\cal H} = {\cal H_K} + {\cal H_V} \label{Ham}
\eea
where 
\bea
{\cal H_V} = \left(
\begin{array}{cc}
{\cal V} & \bzero \\
\bzero & {\cal V}
\end{array}
\right),\;\; {\cal V} = \left( 
\begin{array}{cc}
V_A(\br) & 0 \\
0 & V_B(\br)
\end{array}
\right) \label{Hv}
\eea
and states evolve via the Schr\"odinger equation \prn{Schrod} with ${\cal H}$ instead of ${\cal H_K}$.

\subsection{Review of earlier work: key results}

We now briefly review the results of Brey and
Fretig\cite{brey2006a,brey2006b} for states in a graphene nanoribbon that
are terminated by zig-zag edges in the absence of any external
potentials. Edge states including spin-orbit interaction have also been studied,\cite{sandler2009} but we shall not consider this here. The nanoribbon occupies the region $-\infty < x < \infty$,
$0 \le y \le W$ as shown in \Fig{grscheme}, with the edge at $y=W$
being terminated by $A$ type atoms and that at $y=0$ by $B$ type
atoms. The nanoribbon may also be thought of as a one dimensional
crystal with a basis of atoms as shown in \Fig{grscheme}. The 1d Brillouin zone of this system, shown in \Fig{ebz}, is the segment $\bQ_-\bQ_+$.  Within this scheme, the Dirac point $\bK_-$ is equivalent to the point $\bE_+$ in the 1d Brillouin zone, and similarly $\bK_+$ is equivalent to $\bE_-$.

The problem has translational symmetry in the $x$-direction, and the
energy eigenstates of \prn{HK} are of the form $\Psi_\sigma(x,y) \equiv \psi_{\sigma \alpha
  v}(x,y) = e^{i k x} \phi_{\sigma \alpha v} (y)$, where $k$ is a
wave-vector (momentum) measured from the Dirac points ($\bE_+$,
$\bE_-$).  The boundary conditions\cite{brey2006b} are
\bea
\psi_{\sigma A \pm}(0) = 0; \;\;\;\; \psi_{\sigma B\pm}(W) = 0 \label{FreeBC}.
\eea
The equations for the valleys therefore nicely decouple, i.e., for
zigzag edge nanoribbons, valley index is a good quantum number.

\begin{figure}
\centerline{\epsfxsize=\myfigwidth \epsfclipon \epsfbox{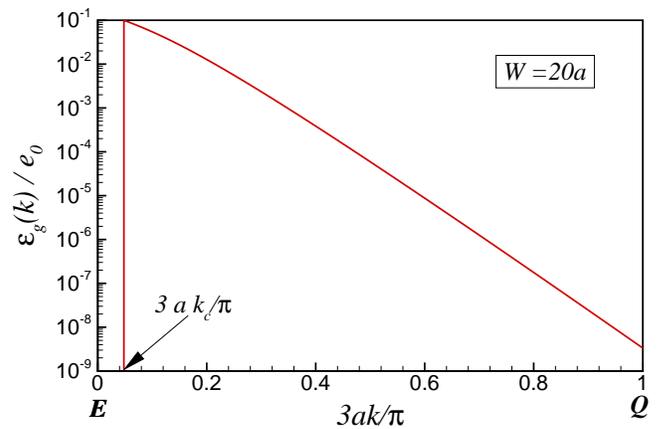}}
\caption{Energy gap of the edge-states as a function of the momentum in EBZ.}
\label{gap}
\end{figure}

We focus here on the edge-states which play a pivotal part in our
work. Edge-states are those eigenstates for which $\phi_{\sigma \alpha
  v}(y)$ are appropriately ``exponentially decaying'' functions, i.e.,
they are wave propagating along the $x$-direction with amplitudes
``localized'' on the edge atoms. Thus, for a given $k$,
\bea
\phi_{\sigma \alpha v}(y) = F_{\sigma \alpha v} e^{q y} + G_{\sigma \alpha v} e^{-q y},
\eea
$q = \sqrt{k^2 - (\varepsilon/a e_0)^2}$ where $\varepsilon$ is the energy eigenvalue, i.e., ${\cal H_K}\Psi_\sigma = \varepsilon  \Psi_\sigma  $, $F$ and $G$ are constants. One finds that $q$ satisfies the equation
\bea
\frac{q}{\tanh{(qW)}} = \zeta k \label{secular}
\eea
where $\zeta = -v$ is $+1$ for $\bE_+$ ($\bK_-$ valley) and $-1$ for $\bE_-$ ($\bK_+$ valley). It is easily concluded that edge-states are possible only for momentum in the ``edge Brillouin zone'' (EBZ, see \Fig{ebz}) $[\bQ_-,\bE_-] \cup [\bE_+, \bQ_+]$, and there are only ``right movers'' at the $\bE_+$ valley and, only ``left movers'' at the $\bE_-$ valley. In particular, there are edge-states around $\bE_+$ only for $k \in [k_c,  \frac{\pi}{3 a}]$, and for $k \in [-\frac{\pi}{3 a}, -k_c]$ around $\bE_-$, where
\bea
k_c = \frac{1}{W} \label{kc}
\eea
Note that in the limit of $W \gg a$, we see that EBZ is about one third of the BZ of the nanoribbon, i.e., roughly, there is one edge-state for every three edge atoms. Now, for each $k$ in the EBZ, there are two edge-states with energy $\pm \varepsilon(k)$. For large $k$ ($\sim \pi/3a$), one obtains the energy dispersion as
\bea
\frac{\varepsilon(k)}{e_0} \approx \pm 2 k a e^{-k W}, \;\;\;\; \frac{\varepsilon_g(k)}{e_0} \approx 4 k a e^{-k W}
\eea
where the $+$ sign is for the ``conduction band'', and the $-$ is for the valance band. One can define a gap function as the difference in the energies of the conduction and valance band for each $k$. It is interesting to note the remarkable {\it exponential distribution} of the energy gap (see, for example, \Fig{gap}); as will become evident, this character of zigzag edge terminated nanoribbons play a crucial role in their nonlinear response. The edge-state wavefunctions, to be used in the discussion below, are
\bea
\phi_{\sigma A -}(y) & = & \sqrt{\frac{2 q}{\sinh{(2 q W)- 2 q W }}} \sinh(q y)\non \\
\label{wfzero}\\
\phi_{\sigma B -}(y) & = & \mp \sqrt{\frac{2 q}{\sinh{(2 q W)- 2 q W }}} \sinh(q(W- y))\non 
\eea
where the $\mp$ stands for valance and conduction bands. These wavefunctions are for the $\bE_+$ ($\bK_-$) valley, with similar expressions for those at $\bE_-$. These analytical results have been tested against full numerical calculations of the tight binding model.\cite{brey2006b}

In addition to the edge-states, the nanoribbon also supports ``bulk states'' and these can be obtained by letting $q$ in \prn{secular} to be imaginary. The key point to be noted is that in a finite width nanoribbon, the energy of the bulk states, in magnitude, are at least $\sim a e_0/W$ -- in this sense the bulk states are ``fully gapped'' for narrow ribbons. The same argument applies to energy differences between different bulk bands, they are of the order of $\sim a e_0/W$.

\section{Graphene Nanoribbons with an Edge Potential}
\label{sec:GNREdgePotential}

The specific problem of interest to us is when the nanoribbon is
subjected to {\it edge potentials}. In terms of the potential operator
${\cal V}$, the edge potential \prn{Hv} can be expressed as
\bea
{\cal V} = \left( 
\begin{array}{cc}
a V_A \delta(y - W) & 0 \\
0 & a V_B \delta(y)
\end{array}
\right)  \label{edgeV}
\eea
where $\delta(\cdot)$ is the Dirac delta function, $V_A$ and $V_B$
have units of energy, and represent, respectively, the values of
constant potentials acting on the edge atoms $A$ $(y=W)$ and $B$ $(y=0)$
(see \Fig{grscheme}). It is also possible to  generalize the
problem to include edge Zeeman fields as will be done later in the
paper.

\subsection{Qualitative Discussion}

We first discuss why such an edge potential \prn{edgeV} produces
highly non-linear Weber-Fechner like responses. Note that the
potential \prn{edgeV} retains the translational symmetry of the
problem in the $x$-direction, i.~e., the momentum $k$ introduced in
the previous section continues to be a good quantum number. Also, the
potential does not mix valley states. Therefore, the eigenstates on
the application of the edge potential will be linear combination of
states with the same momentum $k$. Consider now $|V_A| \sim |V_B| \sim
V$, where $V$ is a characteristic magnitude of the edge potential with
$V \ll a e_0/W \lesssim e_0$. Based on the discussion in the previous
section, it is evident that the applied edge potential will have
little effect on the bulk states due to large energy denominators that
result in the quantum mechanical mixing by the edge potential. Edge-states are however not spared. Some of them are very strongly affected
by the edge potential. In particular states with $k > k_m$ are
strongly affected in that the edge potential mixes the valance and
conduction band edge-states. We can estimate $k_m$ as follows: the matrix element of ${\cal V}$ that mixes the conduction band $\ket{k_{C}^{edge}}$ and valance band states $\ket{k_{V}^{edge}}$ is roughly (using \prn{wfzero})
\bea
\gamma(k) = |\bra{k_{C}^{edge}}{\cal V}\ket{k_{V}^{edge}}| \sim 4 k a V. 
\eea
Strong mixing of states is obtained whenever
\bea
\varepsilon_g(k) \lesssim \gamma(k)\;\; \Longrightarrow \;\; 4 k a e_0 e^{-k W} \lesssim 4 k a V. 
\eea
Thus we find that all the edge-states with $k > k_m$ are strongly affected by the edge potential, with
\bea
k_m = - \frac{1}{W} \ln{\left(\frac{V}{e_0}\right)}, 
\eea
and those with $k < k_m$ are relatively unaffected. Interestingly, we do not expect any response when $V$ is below a threshold $V_{th} = e_0 e^{-\frac{\pi W}{3a}}$.  The physical origin of the Weber-Fechner like responses is now immediate. The response of a system is typically a change in the  charge or spin density from the case with no edge potential. It is clear that the mixing of conduction and valance band states can push the valance band edge-states with $k > k_m$  above (or bring conduction band edge-states below) the chemical potential. The response to the edge potential arises solely from the change in occupation of these states. Other edge-states $k < k_m$ and bulk states are essentially unaffected and do not contribute to the response. These simple arguments suggest that such edge potentials will evoke highly nonlinear response of the Weber-Fechner kind in a zigzag edge terminated graphene nanoribbon. Observe also that $k_m$ depends inversely on the width $W$ of the nanoribbon. Motivated by this discussion, we shall now obtain the edge-state dispersion and wavefunctions in presence of the edge potential of the form \prn{edgeV} to make these ideas precise.

\subsection{Edge-states in presence of an edge potential}
As noted, the edge potential \prn{edgeV} preserves the translational symmetry in the $x$-direction and the energy eigenstates have the form $\psi_{\sigma \alpha -}(x,y) = e^{i k x}\phi_{\sigma \alpha -}(y)$ for the $\bE_+ (\bK_-)$ valley, and satisfy 
\bea
-e_o a (D_y + k) \phi_{\sigma B -}(y) + V_A a \delta(y - W) \phi_{\sigma A-}(y) & = & \varepsilon  \phi_{\sigma A -}(y) \non \\
\label{Aeqn} \\
e_0 a (D_y - k) \phi_{\sigma A -}(y) + V_B a \delta(y) \phi_{\sigma B -}(y) & = & \varepsilon \phi_{\sigma B -}(y) \non \\
\label{Beqn}
\eea
where $\varepsilon$ is the energy eigenvalue and $D_y$ is the derivative with respect to $y$. The equations \prn{Aeqn} and \prn{Beqn} can be solved in the region $0 < y < W$ as
\bea
\phi_{\sigma A -}(y) & = & F e^{q y} + G e^{-q y} \label{phiA} \\
\phi_{\sigma B -}(y) & = & \frac{1}{\varepsilon} \left( F (q -k) e^{q y} - G (q + k) e^{- q y} \right) \label{phiB}
\eea
Integrating \prn{Aeqn} from $W - \eta$ to $W + \eta$ and letting $\eta \rightarrow 0$, we obtain
\bea
e_0 \phi_{\sigma B -}(W) + V_A \phi_{\sigma A -}(W) = 0 \label{topBC}
\eea
and by integrating \prn{Beqn} from $y=-\eta$ to $y= \eta$, we get
\bea
e_0 \phi_{\sigma A -}(0) + V_B \phi_{\sigma B -}(0) = 0 \label{botBC}
\eea
Equations \prn{topBC} and \prn{botBC} provide the necessary boundary conditions. The resulting secular equation for the energy eigenvalue reads
\bea
\left(1 - \frac{V_A V_B}{e_0^2} \right) \frac{q}{\tanh{(q W)}} +  \left(\frac{V_A + V_B}{e_0} \right) \frac{\varepsilon}{a e_0} \non \\
 = \left(1+\frac{V_A V_B}{e_0^2} \right) k 
 \label{secularV}
\eea
The secular equation for the $\bE_- (\bK_+)$ valley will have a $-$ sign on the right hand side.

\subsection{Half Sheet}
\label{sec:hs_ek}

We shall first consider a half sheet of graphene which occupies the region $y \ge 0$ -- the edge is terminated by $B$ type atoms (see \Fig{grscheme}) and a potential $V (\ll e_0)$ acts on these edge atoms. The secular equation  \prn{secularV} can be readily solved to obtain the dispersion
\bea
\varepsilon(k) = \frac{2 e_0 V}{e_0^2 + V^2} a k \approx 2 \frac{V}{e_0} a k, \label{hsdisp}
\eea
near $\bE_+$ (a negative sign is obtained for states near $\bE_-$).
A constant edge potential on a half sheet makes the edge modes {\it linearly dispersing}, i.~e., moving with a constant velocity which is proportional to the applied potential. As anticipated in the qualitative discussion all the edge states are affected (since $k_m \approx 0$). Moreover, these edge-states are such that, for a positive edge potential $V$, the states near $\bE_+$ are right movers, and those at $\bE_-$ are left movers. Most interestingly, by applying a negative edge potential, the sign of the velocity can be reversed, i.~e., states near $\bE_+$ become left movers etc.!
 The edge-state wavefunctions (at $\bE_+$) are
\bea
\phi_{\sigma B -}(y) = \sqrt{\frac{2 q}{(1 + (V/e_0)^2)}} e^{-q y},\;\; \phi_{\sigma A -}(y) = -V \phi_{\sigma B -}(y) \non \\
\label{hsWFns}
\eea
where $q = \frac{e_0^2 - V^2}{e_0^2 + V^2} k$.

It must be noted that for the half sheet, the bulk states near $\bE_+$ are also affected by the edge potential. However, the bulk dispersion (and gap) is not exponential as in the case of edge states. Therefore, only small corrections to the bulk states are expected in a small region near the Dirac points.
 
We now proceed to consider the case of finite width nanoribbons. We shall consider two special cases of edge potentials. First is a symmetric edge potential where $V_A = V_B = V$ and second is an anti-symmetric edge potential $V_A = -V_B = V$.

\subsection{Nanoribbons with Symmetric Edge Potential}
\label{sec:sym_ek}

\begin{figure}
\centerline{\epsfxsize=\myfigwidth \epsfclipon \epsfbox{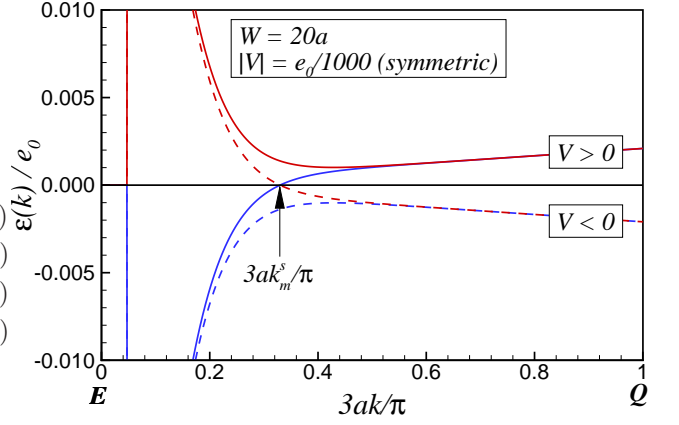}}
\caption{(color online) Dispersion of edge states in the presence of a symmetric edge potential. Blue line is the valance band and red line the conduction band. Solid lines are for a positive value of the edge potential, while the dashed ones are for a negative value of the edge potential. Note the near linear dispersion of the edge states near $\bQ$.}
\label{sym_ek}
\end{figure}

A symmetric edge potential $V_A = V_B = V$, shifts the energies of the conduction and valance bands in the same direction, i.~e, for every $k$, the energies of both the bands either increase or decrease depending on the sign of $V$.
For such a potential with $V_A = V_B = V$, the secular equation \prn{secularV} reduces to 
\bea
\left(1 - \frac{V^2}{e_0^2} \right) \frac{q}{\tanh{(q W)}} +  2\left(\frac{V}{e_0} \right) \frac{\varepsilon}{a e_0} = \left(1+\frac{V^2}{e_0^2} \right) k \label{secularS}
\eea
where $\varepsilon$, the energy eigenvalue, is one of $\pm e_0 a \sqrt{k^2 - q^2}$. The two solutions evolve from the conduction (valance) bands of the zero edge potential problem. For the case of the valance band, we see that edge state solutions appear only when $k > k_c^V$, where
\bea
k_c^V = \frac{e_0^2 - V^2}{(e_0 + V)^2} \frac{1}{W}
\eea
and similarly for the conduction band only when $k > k_c^C$ with
\bea
k_c^C = \frac{e_0^2 - V^2}{(e_0 - V)^2} \frac{1}{W}
\eea
For edge potentials with $V \ll e_0$, $k_c^V$ and $k_c^C$ are only very slightly different from the case without the edge potential \prn{kc} and do not affect the physics in any significant way.
When $k$ is near $k_c$, the dispersion is essentially unaffected by the presence of a small edge potential. However, for $k$s close to the EBZ boundary, $k \approx \frac{\pi}{3a}$, the dispersion for both the conduction and valance bands is
\bea
\frac{\varepsilon(k)}{e_0} \approx   \frac{2V}{e_0}  a k
\eea
much like the case of the half sheet discussed in the previous section. \Fig{sym_ek} shows both the bands in the EBZ obtained from the numerical solution of \prn{secularS}; they have all the features anticipated in the qualitative discussion above -- states near the EBZ center (near $\bE$) are unaffected while those near EBZ boundary (near $\bQ$) are strongly affected. An important feature is that for a positive (negative) edge potential the valance (conduction) band crosses the zero energy, and the conduction (valance) band has a minimum (maximum). It is clear that the wave vector $k_m^s$  (the superscript $s$ indicates that we are dealing with a symmetric edge potential) at which the valance (conduction) band crosses the zero of energy has a key role to play in the response. This wavevector $k_m^s$ can be immediately obtained from \prn{secularS}:
\bea
k_m^s = - \frac{1}{W} \ln{\left( \frac{V}{e_0} \right)}
\eea
Near $k_m$, the band that crosses the zero of energy disperses linearly with a velocity $v_m$ given by
\bea
\frac{v_m}{v_F} =  - \frac{2 e_0 V}{e_0^2 - V^2} \ln{\left(\frac{|V|}{e_0} \right)} \label{eqn:sym_vel}
\eea
which we see is independent of the ribbon width.

Due to the symmetry of the edge potential the wavefunctions are such that the total weight on the $A$ sublattice is equal to the total weight on the $B$ sublattice.

\subsection{Nanoribbons with Anti-Symmetric Edge Potential}
\label{sec:asym_ek}

\begin{figure}
\centerline{\epsfxsize=\myfigwidth \epsfclipon \epsfbox{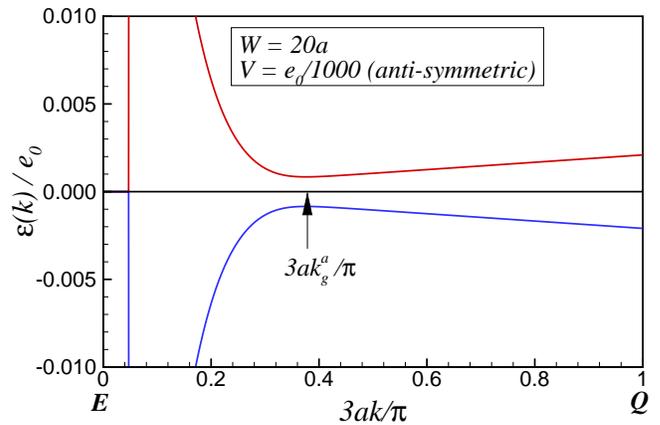}}
\caption{(color online) Dispersion of edge states in the presence of an anti-symmetric edge potential. Blue line is the valance band and red line the conduction band.  Note the near linear dispersion of the edge states near $\bQ$; also, the lowest energy gap obtains at $k_g^a$.}
\label{asym_ek}
\end{figure}

In the second case, we consider an antisymmetric edge potential such that $V_A = -V_B = V$. This edge potential is such that it mixes the conduction and valance band states, with a resulting ``level repulsion''. The secular equation \prn{secularV} now becomes
\bea
\left(1 + \frac{V^2}{e_0^2} \right) \frac{q}{\tanh{(q W)}} = \left(1-\frac{V^2}{e_0^2} \right) k \label{secularA}
\eea
for states near $\bE_+$ (a negative sign on the r.~h.~s.~ will give the secular equation for states at $\bE_-$). Moreover, in this case, energy eigenvalues come in $\pm$ pairs due to the (anti-)symmetry of the potential. We find that edge-states appear whenever $k \ge k_c^a$,
\bea
k_c = \frac{e_0^2 + V^2}{e_0^2 - V^2} \frac{1}{W}
\eea
Again, we find that that states with $k$ near $k_c$ are unaffected by a small edge potential, while those near the EBZ boundary are strongly affected and disperse linearly as
\bea
\frac{\varepsilon(k)}{e_0} \approx \mp  \frac{2V}{e_0} a k
\eea
for the valance and conduction bands respectively. \Fig{asym_ek} shows the dispersion of the valance and conduction bands obtained from the solution of \prn{secularA}. The smallest gap is no longer at the boundary of the EBZ, but occurs at a wavevector $k_g^a$ which is calculated as
\bea
k_g^a \approx \frac{1}{W} \left(- \ln{\left(\frac{|V|}{e_0}\right)} + \half \ln{\left(- \ln{\left(\frac{|V|}{e_0}\right)} \right)} \right)
\eea
and the energy gap is
\bea
\frac{\varepsilon_g}{e_0} \approx -\frac{2a}{W}  \frac{|V|}{e_0} \ln{\left(\frac{|V|}{e_0}\right)}. \label{eqn:asym_gap}
\eea
with, again, a  strongly non linear dependence on the applied edge potential.
The wavefunctions of this problem have qualitatively different features from the case of the symmetric potential. The wavefunctions are relatively unaffected for states with $k \approx k_c^a$ in that the weight in the $A$ sublattice and $B$ sublattice are about equal. However, assuming $V > 0$,  most of the weight is in the $B$ ($A$) sublattice for valance (conduction) band states with  $k \gtrsim k_m^a$. In fact the difference of weights is a monotonic function of $k$ (see \Fig{rhok}, to be discussed in detail later).

Armed with the discussion of the eigenstates of the system in presence of edge potentials we now proceed to discuss the responses. Results similar to those in \fig{sym_ek} and \fig{asym_ek} have been obtained by Wimmer\cite{WimmerThesis}.

\section{Density Response with Edge Potentials}
\label{sec:dens_resp}

This section focusses on the zero temperature density response of zigzag edge terminated graphene systems that are subjected to an edge potential. Throughout this section, we shall assume that the chemical potential of the system is maintained at zero. We shall discuss separately the three cases that were discussed in the previous section.

\subsection{Half Sheet}
\label{sec:hs_denresp}

From the dispersion for the edge-states \prn{hsdisp}, it is immediate that for a negative edge potential, the edge states are pushed below the zero chemical potential. For a small edge potential, there is no further change of occupancy, and therefore there is no density response. However, for a positive edge potential, the edge states are thrown above the chemical potential, and therefore every edge-state in the EBZ is left unoccupied. It is evident, therefore, change in number of electrons (per repeat unit in the $x$ direction)  is
\bea
\Delta N = \Delta N_{\uparrow} + \Delta N_{\downarrow} = - \frac{2}{3} \Theta(V)
\eea
where $\Theta(\cdot)$ is the Heaviside step function, the factor $3$ in the denominator arises from the fact that the EBZ is a third of the full 1D BZ, and the factor $2$ in the numerator from two spins.  It is also interesting to study how this loss of particles is distributed on the half sheet:
\bea
\Delta n_A(y) & = & -\Theta(V) \frac{V^2}{(e_0^2 - V^2)} \frac{a}{\pi y^2} \left(1 - e^{-\frac{2 \pi \nu y }{3 a}} (1 +\frac{2 \pi \nu y }{3 a} ) \right) \non\\
\label{hs_cdens}\\
\Delta n_B(y) & = & -\Theta(V) \frac{e_0^2}{(e_0^2 - V^2)} \frac{a}{\pi y^2} \left(1 - e^{-\frac{2 \pi \nu y }{3 a}} (1 +\frac{2 \pi \nu y }{3 a} ) \right) \non \\
\eea
where $\nu = \frac{(e_0^2 - V^2)}{(e_0^2 + V^2)}$. It is interesting to note that $n_A(y) + n_B(y)$ has an integrable power law decay, with  diverging higher moments.

\subsection{Nanoribbons with Symmetric Edge Potential}
\label{sec:sym_denresp}

The change in the number of particles in the case can be calculated as follows. Note from \Fig{sym_ek} that the occupancy of all states with $k < k_m^s$ is unaffected by the edge potential. For the case with $V > 0$ ($V < 0$) and $k > k_m^s$, the valance (conduction) band states are lifted above (pushed below)the chemical potential and are hence unoccupied (occupied). One finds that the change in the number of particles per repeat unit is
\bea
\Delta N = \Delta N_\uparrow + \Delta N_\downarrow = -2 \signum{(V)}  \left( \frac{1}{3} + \frac{a}{\pi W} \ln{\left(\frac{|V|}{e_0}\right)} \right) \non \\
\label{eqn:sym_res}
\eea
which has a strongly nonlinear Weber-Fechner like behaviour. It must also be noted that, the response will vanish for very small edge potentials. The threshold potential for a nonzero response is obtained as
\bea
V_{th} = e_0 e^{-\frac{\pi W}{3 a}} \label{eqn:threshold}.
\eea
We can rewrite \prn{eqn:sym_res} as
\bea
\Delta N = -2 \signum{(V)} \frac{a}{\pi W} \ln{ \left(\frac{|V|}{V_{th}} \right)},
\eea
which is in the standard Weber-Fechner form.

\subsection{Nanoribbons with Anti-symmetric Potential}
\label{sec:asym_denresp}

\begin{figure}
\centerline{\epsfxsize=\myfigwidth \epsfclipon \epsfbox{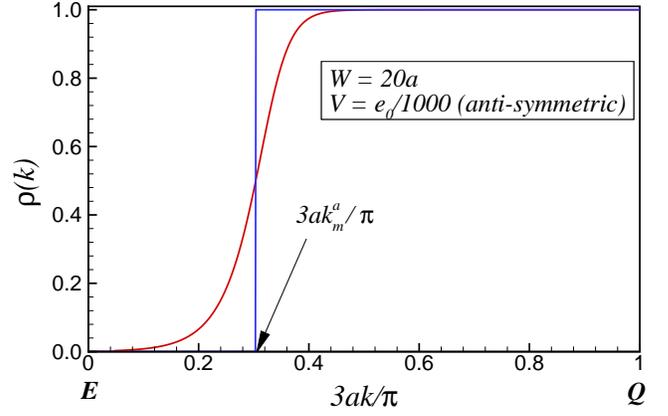}}
\caption{(color online) Contribution to the ``charge density wave'' amplitude from edge modes. The total ``charge density wave'' amplitude $C$ is well approximated by $\rho_{approx}(k)= \Theta(k - k_m^a)$ shown by the blue line; the red line is the exact result. }
\label{rhok}
\end{figure}

For an anti-symmetric edge potential, it is immediate that net
particle number response vanishes for any $V$ since the occupation
numbers of the bands are unaffected. However, as noted in section
\ref{sec:asym_ek}, the electronic states are such that the weight on
the $A$ sublattice is different from the $B$ sublattice resulting in a ``density wave'' like response. The density wave response associated with an edge-state of momentum $k$ is 
\bea
\rho(k) = \int_{0}^{W} \D{y} \left( |\phi_{\sigma B -}(y)|^2 - |\phi_{\sigma A -}(y)|^2   \right)
\eea
and this is a monotonic function of $k$. \Fig{rhok} shows a plot of $\rho(k)$ as a function of $k$ for a particular $V >0$. Now the ``density wave order parameter'' $D$ is
\bea
D = \frac{2a}{\pi} \int_0^{\pi/3 a} \D{k} \rho(k).
\eea
A detailed analysis shows that $D$ can be very accurately approximated by taking
\bea
\rho(k) \rightarrow \rho_a(k) = \Theta(k - k_m^a) 
\eea
where
\bea
k_m^a = -\frac{1}{W} \ln{\left(\frac{\sqrt{3} |V|}{e_0}\right)}
\eea
resulting in
\bea
D = 2 \signum(V) \left( \frac{1}{3} + \frac{a}{\pi W} \ln{\left(\frac{\sqrt{3}|V|}{e_0}\right)} \right). \label{eqn:asym_res}
\eea
whenever $|V|$ is greater than an appropriate threshold as in \prn{eqn:threshold}.
For a positive edge potential, the ``density wave'' that arises will
be such that density is higher on the edge $B$, than on edge $A$,
i.~e., a build up of particles at the $B$ edge at the expense of the
$A$ edge. This observation will be important in understanding the spin
response of zigzag edge terminated graphene edge systems as discussed
in the next section.

\section{Spin Response with Edge Zeeman Fields}
\label{sec:spin_resp}

In this section we will explore spin response of zigzag terminated graphene structures to applied edge Zeeman potentials. The edge Zeeman potential operator is 
\bea
{\cal V}_{H\sigma} = \left( 
\begin{array}{cc}
-\sigma a H_A\delta(y - W) & 0 \\
0 & -\sigma a H_B \delta(y)
\end{array}
\right) \label{HvZeeman}
\eea
where $H_A$ and $H_B$ are Zeeman energies due to appropriate fields on edges $A$ and $B$ respectively. As noted in the introduction, such edge Zeeman potentials are encountered in various mean field theories of graphene systems which treat interactions between electrons. We shall now use the results of the previous sections to obtain results for spin response in various cases.

\subsection{Half Sheet}
For the case of the half sheet with a Zeeman potential $H_B = H$, we find the dispersion \prn{hsdisp} as
\bea
\frac{\varepsilon_\sigma(k)}{e_0} = \frac{-2 \sigma e_0 H}{e_0^2 + H^2} a k \approx -2 \frac{\sigma H}{e_0} a k. \label{hsZdisp}
\eea
The spin response can be obtained by considering the filled states, and we obtain the magnetization per repeat unit in the $x$ direction as
\bea
M & = &  \Delta N_\uparrow - \Delta N_\downarrow = -\frac{1}{3} \Theta(-H) - (- \frac{1}{3} \Theta(H)) \non \\ 
&  = & \frac{1}{3} \signum{(H)}
\eea
Not unexpectedly, even an infinitesimal Zeeman field produces a large magnetic response with a magnetization of $1/3$. The spin density, again, dies as a power law similar to \prn{hs_cdens}, with a larger magnitude on the $B$ sublattice atoms and a smaller one of opposite sign on the $A$ sublattice.

\subsection{Nanoribbon with Symmetric Zeeman Potential}
For this case $H_A = H_B = H$, one can show by a particle hole transformation $c^\dagger_{\sigma \alpha} \rightarrow (-)^{\alpha}c_{\bar{\sigma} \alpha}$ ($\alpha$ is $\pm 1$ for $\alpha = A, B$) , the chemical potential is zero when the total number of electrons is maintained at one electron per site. We have  a situation where the total number electrons do not change, but the number of electrons of each spin species changes to produce a magnetic response. We find that the magnetization of the system is (see \prn{eqn:sym_res})
\bea
M = \Delta N_\uparrow - \Delta N_\downarrow = 2 \signum{(H)}  \left( \frac{1}{3} + \frac{a}{\pi W} \ln{\left(\frac{|H|}{e_0}\right)} \right), \label{eqn:sym_M}
\eea
for a Zeeman field larger than a threshold $H_{th}$ given in \prn{eqn:threshold}; again a Weber-Fechner like behaviour. Since the wavefunction of the edge-states have a significant amount of weight at the edges, we find that such a Zeeman edge-potential produces a magnetization that is primarily at the edges. In the present case the magnetizations of both the edges are parallel -- a ``ferro'' configuration - with magnitude roughly given by half of \prn{eqn:sym_M}. The band structure that emerges in the presence of the symmetric potential is such that, at the chemical potential, the dispersion is linear for both spins. However they disperse with opposite velocities given by \prn{eqn:sym_vel} with $V$ replaced by $H$. The direction of the velocities are reversed at the other valley.

\subsection{Nanoribbon with Antisymmetric Zeeman Potential}

For the antisymmetric Zeeman potential, the chemical potential again is zero (one electron per site). Since the total field is zero, there is neither a total density or total spin response. Rather, such a Zeeman field produces a ``spin density wave'' like response where the magnetization on the $A$ sublattice is opposite of that on the $B$ sublattice. For a Zeeman field larger than an appropriate threshold (see \prn{eqn:threshold}) the total magnetization on each sublattice (per unit cell) is 
\bea
M_{B} = - M_A =  \signum{(H)}  \left( \frac{1}{3} + \frac{a}{\pi W} \ln{\left(\frac{\sqrt{3}|H|}{e_0}\right)} \right) \label{eqn:asym_M}
\eea
with a Weber-Fechner like dependence on the edge magnetic field (Zeeman energy). The resulting configuration is such that most of the spin polarization of each sublattice resides at the edges resulting in an edge magnetized state with an ``anti-ferro'' orientation of the edge moments. The system is is gapped with a gap given by \prn{eqn:asym_gap} with $V$ replaced by $H$.

It is particularly interesting to note that for a given magnitude $H$ of the edge Zeeman potential, 
the magnitude of the edge moment for the anti-symmetric potential is larger than that for the symmetric potential. This is key to understanding the magnetic properties of nanoribbons when interactions are included\cite{SomnathThesis}.

\subsection{Doped Nanoribbons with Edge Zeeman Fields}

It is interesting to investigate the properties of doped nanoribbons, particularly with a view towards possibility of application in electronic devices.

\subsubsection{Symmetric Zeeman Potential}
We consider a zigzag edge terminated graphene nanoribbon with a small doping $\delta$ per edge atom ($\delta$ is a dimensionless number). A positive $\delta$ corresponds to hole doping, and  negative $\delta$ to electron doping. 
\begin{widetext}
As noted earlier, the dispersion near the zero energy (i.e., near $k \approx k_m$) is given by
\bea
\frac{\varepsilon_\sigma(k)}{e_0} =-\sigma \left(\frac{v_m}{v_F} a(k - k_m) - \left( \frac{W}{a} \right) s_m a^2(k - k_m)^2\right)
\eea
where $v_m$ is as defined in \prn{eqn:sym_vel} with $s_m$ 
\bea
s_m = - \frac{e_0 H}{e_0^2-H^2} \left( \frac{e_0^2 + H^2}{e_0^2 - H^2} \ln{\left(\frac{|H|}{e_0} \right)} + 2 -  \frac{4 H^2}{e_0^2 - H^2} \ln{\left(\frac{|H|}{e_0} \right)}^2\right)
\eea
\end{widetext}
is a dimensionless number that depends on the potential.
Using this dispersion relation, we can calculate the change in the magnetization per edge atom as
\bea
\Delta M (\delta) = -2 \pi \left(\frac{v_F}{v_m} \right) s_m  \left(\frac{W}{a} \right) \delta^2 \label{eqn:sym_deltaM}
\eea
to lowest order in $\delta$. We see that the edge magnetization falls quadratically with doping with coefficient proportional to the width of the ribbon. The change in energy (per repeat unit in the $x$-direction) as a function of doping is
\bea
\frac{\Delta E^S(\delta)}{e_0} = \frac{\pi v_m}{v_F} \delta^2. \label{eqn:sym_deltaE}
\eea

\subsubsection{Antisymmetric Zeeman Potential}

In this case the magnetization per edge atom falls as
\bea
\Delta M = -|\delta| \label{eqn:asym_deltaM}
\eea
when the doping per edge atom is $\delta$. Energy changes according to
\bea
\frac{\Delta E^A(\delta)}{e_0} =  2 \frac{\varepsilon_g}{e_0} |\delta|.  \label{eqn:asym_deltaE}
\eea
where $\varepsilon_g$ is the energy gap given in \prn{eqn:asym_gap}.

\section{Comparison with Full Tight Binding Simulations}
\label{sec:compare_tb}

In this section we present comparison and validation of the theoretical results of the previous sections with full tight binding calculations of undoped and doped systems. We discuss only the case with the edge Zeeman fields. The tight binding calculations are performed using the momentum basis to represent \prn{HTB} with appropriately applied edge potentials. We have used up to 10000 $k$-points in the BZ $\bQ_-\bQ_+$ (see \Fig{ebz}) in the calculations reported here.

\begin{figure}
\centerline{\epsfxsize=\myfigwidth \epsfclipon \epsfbox{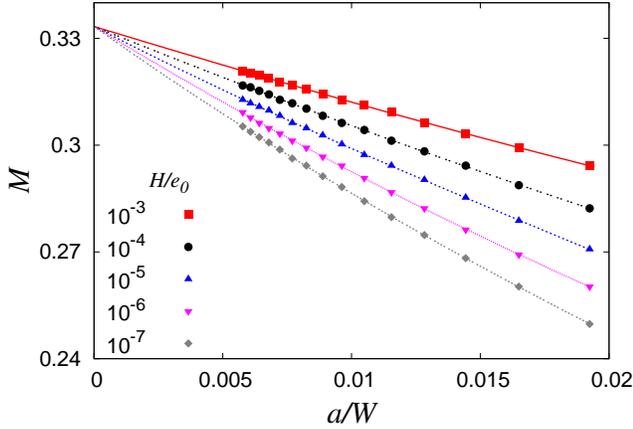}}
\caption{Dependence of edge magnetization $M$ on the width $W$ of the undoped graphene nanoribbon for different symmetric edge Zeeman fields. Points are results of the tight binding calculations, lines are fits to theory.}
\label{symM}
\end{figure}

\Fig{symM} shows a plot of the calculated moment $M$ per edge using the tight binding calculations as a function of the width $W$ of the nanoribbon with a symmetric edge Zeeman field $H$. Results are shown  for five orders of magnitude for the Zeeman field $H$. We find that the Weber-Fechner behaviour predicted in \prn{eqn:sym_M} is well reproduced in the tight-binding calculations. In particular, the coefficient of the $(a/W)$ term in \prn{eqn:sym_M} is reproduced to with in a percent.

\begin{figure}
\centerline{\epsfxsize=\myfigwidth \epsfclipon \epsfbox{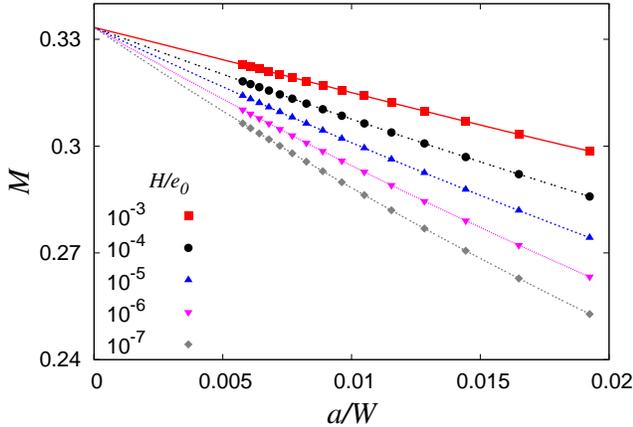}}
\caption{Dependence of edge magnetization $M$ on the width $W$ of the undoped graphene nanoribbon for different anti-symmetric edge Zeeman fields. Points are results of the tight binding calculations, lines are fits to theory.
}
\label{asymM}
\end{figure}

Results of a similar study of nanoribbons with an antisymmetric Zeeman potential is shown in \Fig{asymM}, where again we see that  Weber-Fechner law predicted \prn{eqn:asym_M} is realized. For a given $H$ and $W$, we see that the antisymmetric potential produces a larger edge moment, as predicted in \prn{eqn:sym_M} and \prn{eqn:asym_M}. 

We have also investigated various other quantities (e.g., the gap \prn{eqn:asym_gap}) using the tight binding calculations and compared them with our analytical predictions and found excellent agreement in all cases\cite{SomnathThesis}.

\begin{figure}
\centerline{\epsfxsize=\myfigwidth \epsfclipon \epsfbox{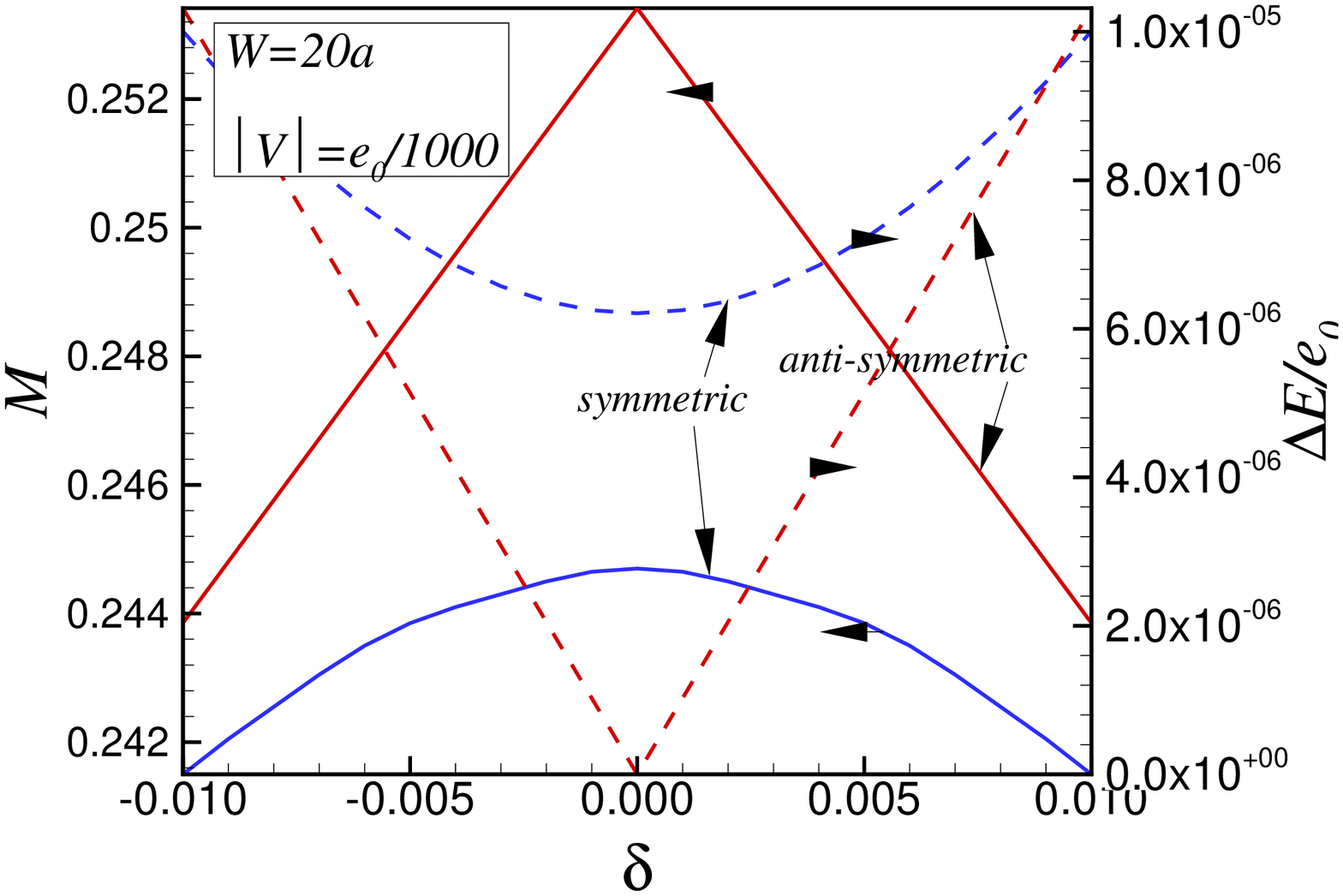}}
\caption{(color online) Dependence of magnetization $M$ and change in energy $\Delta E$ (referred to the undoped anti-symmetric case) as a function of doping $\delta$ per edge atom. Blue lines correspond to the symmetric case, while the red lines represent the anti-symmetric case. Both $M$ and $\Delta$ are quadratic function of the doping $\delta$ for the symmetric case, while they are linear for the anti-symmetric case.}
\label{delMe}
\end{figure}

The dependence of the magnetization and energy on the doping $\delta$ obtained from tight binding calculations is shown in \fig{delMe}. The results \prn{eqn:sym_deltaM}, \prn{eqn:sym_deltaE}, \prn{eqn:asym_deltaM} and \prn{eqn:asym_deltaE} are recovered from the tight binding calculations both for hole and electron doped systems.

\section{Summary and Conclusion}
\label{sec:summary}

In summary, we have shown that zigzag edge terminated graphene
nanoribbons show highly nonlinear Weber-Fechner like response to static edge
potentials. These results were obtained from analytical considerations
of the continuum Dirac equations for graphene. These analytical
results were validated qualitatively and quantitatively with full
tight binding calculations. It is not often that such a highly
nonlinear response of even a noninteracting system has been obtained
analytically, and this is a rare example of a quantum system that shows Weber-Fechner like response found in sensory organs like the eye and ear. It is also interesting to compare graphene nanoribbons to the well known example of a $p-n$ diode (Schockley equation)\cite{Ashcroft1976} where the diode current depends exponentially  on the voltage divided by the thermal voltage $k_BT/e$ ($k_B$--Boltzmann constant, $e$ electron charge).  In the present case, the Weber-Fechner response is an intrinsic zero-temperature property of the zigzag edge graphene nanoribbon.

Our work also suggests many possible uses of graphene nanoribbons with
applied edge potentials. In the case of the symmetric Zeeman
potentials, one obtains a system with nearly linearly dispersing
particles in 1D. The velocity of these particles can be controlled by
tuning the edge potential and the width of the
nanoribbons. Antisymmetric edge potentials provide a route to produce
tunable 1D gapped systems. Furthermore antisymmetric Zeeman potentials
produce half-metallic edges.\cite{son2006,young2006} Clearly, these are fruitful
avenues for further investigation. It will be particularly interesting to realize such a physical system using cold atom optical lattices\cite{Giorgini2008,Bloch2008} where the kinetic energy, interactions, and, hopefully soon, temperature can be tuned.

It will be interesting to investigate how physical effects not
considered in detail in this work affect the nonlinear
response. Primary among them are the second neighbor
hopping\cite{Neto2006} $t'$, temperature effects, disorder and
interactions. In particular, the Weber-Fechner like behaviour will
require very low temperatures (compared to the energy $a e_0/W$). A simple estimate for the temperature required to observe Weber-Fechner like response can be obtained by comparing the gap scale in \prn{eqn:asym_gap} to the temperature. For a ribbon of width $\sim 10 a$,  Weber-Fechner response should be observable  for edge potentials of order $10^{-2}e_0 - 10^{-4}  e_0$ for temperatures below $\sim 10^{-2}e_0 - 10^{-4} e_0$. Taking $e_0 \sim 2$eV, we estimate this temperature to be below 10K. 

The
Weber-Fechner like response is likely to be protected against weak
potential disorder due to the chiral nature of the edge states. The most interesting effects 
arise due to the presence of interactions among the electrons and associated Luttinger liquid effects.\cite{Hausler2008} In this case, the edge potentials can be ``self generated'' and can lead to interesting magnetic properties of the system. A detailed study of these effects has been completed\cite{SomnathThesis} and will be reported elsewhere. This study also suggests that Weber-Fechner like response is obtained even when the edge potential is spread over three/four rows of atoms in proximity of the edge.

\subsection*{Acknowledgements}
We acknowledge support of this work by DST, India through MONAMI and Ramanujan grants. VBS thanks Jayantha Vyasanakere for an illuminating discussion on the Weber-Fechner law. The authors are grateful to Ganapati Sahoo and Rahul Pandit for computational resources.

\bibliography{ref}

\end{document}